\documentstyle[floats,multicol,aps,epsf,epsfig,prb]{revtex}
\newcommand \bea {\begin{eqnarray} }
\newcommand \eea {\end{eqnarray}}

\begin{document}
\draft
\twocolumn[\hsize\textwidth\columnwidth\hsize\csname @twocolumnfalse\endcsname
\title{ 
Bound-State Instability of the Chiral Luttinger Liquid in One-Dimension. 
}
\author{ A. F. Ho$^{1}$ and  P. Coleman$^{1}$ }
\address{
$^1$Serin Laboratory of Physics, Rutgers University, 136 Frelinghuysen Rd.,
Piscataway, New Jersey 08854 USA}
\maketitle
\date{\today}
\maketitle
\begin{abstract}  

We have developed a new boot-strap method for solving a  class of interacting
one-dimensional chiral fermions.   The conventional model
for interacting right-moving electrons  with spin 
has an $SO(4)$ symmetry, and can be written as four
interacting Majorana fermions, each with the same velocity.
We have found a method for solving some cases when the
velocities of these Majorana fermions are no longer equal.
We demonstrate in some detail the remarkable result that corrections to 
the non skeleton self-energy identically vanish for these 
models, and
this enables us to solve them exactly.  For the cases
where the model can be solved by bosonization,
our method can be explicitly checked. However, we are also
able to solve some new cases where the excitation spectrum
differs qualitatively from a Luttinger liquid. 

Of particular interest, is the so-called $SO(3)$ model, 
where a triplet of Majorana fermions moving at one velocity,
interact with a single Majorana fermion moving at another
velocity.  We show using our method, that a sharp bound 
(or anti-bound) state  splits off
from the original Luttinger liquid continuum,
cutting off the X-ray singularity to form a broad incoherent excitation
with a lifetime that grows linearly with frequency. 

\end{abstract}
\vskip2pc]

\section{INTRODUCTION}

The anomalous normal state behavior discovered in the cuprate 
superconductors has stimulated enormous interest in the possibility
of new kinds of electronic fluid that might provide an alternative
to Fermi liquid behavior.
The classic model for non-Fermi liquid behavior is provided by
the one-dimensional electron gas, 
where  the generic fixed point behavior is 
a Luttinger  Liquid.\cite{Haldane} Thanks to a wide array of
non-perturbative techniques, there is a rather solid understanding
of the Non-Fermi Liquid properties in such $1d$ systems.
Motivated by an early suggestion of Anderson,\cite{PWALutt2}
many authors have attempted to generalize the Luttinger liquid 
concept to higher dimensions.\cite{Metzner,Houghton,Fradkin}  

The Luttinger Liquid in $1d$ is truly special
in that it has no quasi-particle poles but a branch cut
singularity; its correlation functions are  scale-invariant,
with an associated beta function that is zero to all orders in perturbation
theory\cite{Metzner} for a wide range in the coupling:
\[
\beta (g) = 0  .
\]
That the beta function is zero is not in itself 
special to the Luttinger Liquid. 
For example, in the absence of nesting, or a Cooper instability,
the beta function associated with Landau's Fermi Liquid fixed point
is also zero for the forward scattering channel 
\cite{Galivoti,Shankar}

The profound differences between the Luttinger liquid and Landau Fermi
liquid fixed points originate in the special kinematics of one dimension.
In $1d$, the  Fermi surface consists
of just two points $\pm k_f$ where the electrons 
interact very strongly, and asymptotically near these
Fermi points, energy  and momentum conservation  impose a {\sl single}
constraint on scattering processes, 
giving rise to a qualitative enhancement in scattering phase space. 
This causes the electron to lose its  
eigen-state status to the  collective spin and charge density bosonic modes.
Luttinger liquid behavior requires the 
absence of umklapp interactions, and in this case,
left- and right-moving particles are separately conserved.
The spin and charge current densities of the right (or left)
moving particles 
are then simply proportional to the corresponding
spin and charge densities: 
\bea
J_{c}^R = v_{c} \rho_{c}^R,\nonumber
\eea
\bea
J_{s}^R = v_{s} \rho_{s}^R,\nonumber
\eea  
so that the continuity equation assumes a special form
\[
(\partial_{\tau} - i v_{s,c} \partial_x) \rho_{s,c}^R = 0
\] 
As noted long ago
by Dzyaloshinskii and Larkin\cite{D&L,Metzner}, these conservation laws
lead to the vanishing of the $N$-point connected current correlation
functions for $N>2$ (``Loop Cancellation Theorem'', see Section IV), which
leads to a Gaussian theory for the spin and charge bosons in the Tomonaga
Luttinger model, and also for the low energy effective theory of the Hubbard
model in $1d$.

Unfortunately,
the special kinematics of one dimension do not 
survive in higher dimensions, and largely for this reason,
attempts to generalize the Luttinger
Liquid to $d \geq 2$ with strictly local interactions
have been unsuccessful. 
In one dimension, energy and momentum conservation impose
a single constraint on the forward scattering processes, whereas in 
higher dimensions, they impose 
independent constraints on the scattering processes.
These  additional constraints 
eliminate many of the potentially dangerous singularities
present in one dimensional scattering processes, stabilizing the Fermi liquid
in two or higher dimensions.\cite{Metzner,Shankar}
Lin et. al.\cite{Fisher} 
arrived at the same conclusion,
making the passage from one, to two dimensions
by coupling $N$ 
Hubbard chains together and taking the limit $N\rightarrow \infty$.  Whilst
it is possible to circumvent the Fermi liquid in two dimensions
by intrducing 
long-range or singular interactions\cite{Wen,Stamp,PWALutt2},
a route to non-Fermi liquid behavior in $2d$ 
that
involves strictly local interactions has not yet been found.

An alternative approach has however
been advocated by Anderson,\cite{PWALutt1}
who noted that higher dimensional
non-Fermi liquid behavior might derive from the formation of
bound, or anti-bound states above and below the single-particle continuum.
Such bound-states play an important
role in the formation of the one dimensional Luttinger liquid,
where they give rise a finite scattering phase shift at the Fermi energy,
driving the formation of X-ray singularities in the spinon-holon continuum.  

\begin{figure}
\unitlength1.0cm
\begin{picture}(9,7)
\epsfxsize=9.0cm
\epsfysize=7cm
\centerline{\epsfbox{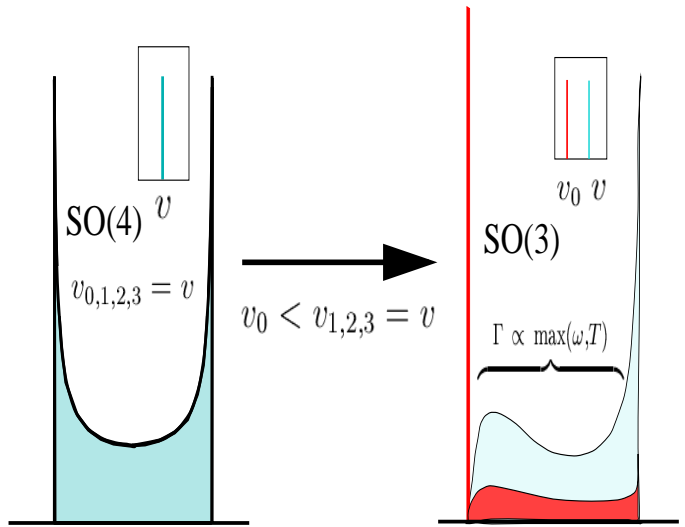}}
\end{picture}
\vskip 0.1 truein
\caption{Schematic diagram showing the evolution of the spectral
weight as we introduce velocity difference to the fermions.
Inset indicates the bare spectral function, without interactions.
\label{schematic}}
\end{figure}

In this paper, we are motivated by this discussion
to examine whether such singularities
are robust against the removal of some of the special
kinematic symmetries of one dimension.
By modifying  the $1d$ kinematics, we show that
it is possible to actually split-off bound-states from the spinon-holon
continuum giving rise to a new type of one-dimensional non-Fermi liquid
that does not rely on the special 1d symmetries mentioned above. 
The key to our idea is as follows. The electron fluid on the Fermi
surface is made up of spin-up and down electrons and holes. Borrowing
from the Dirac equation, we can rewrite the electrons and holes
as charge-conjugation eigen-states:
\[
 c_{\uparrow}=\frac{1}{\surd 2} (\Psi^{(1)} - i \Psi^{(2)}), \qquad
c_{\downarrow}= - \frac{1}{\surd 2} (\Psi^{(3)} + i \Psi^{(0)}) .
\] 
where the $\Psi^{(a)}$ ($a=(0,1,2,3)$ ) represent four chiral Majorana  
fermions\cite{Majorana} such that  $\Psi^{(a)}(x)=\Psi^{(a) \dagger}(x)$.
Instead of changing the interaction, we
modify the scattering kinematics by making one of the 
Majorana fermions to 
have a different velocity
to the others. In the classic Tomonaga Luttinger model, all
four Majorana fermions have the same velocity, (exhibiting the full $SO(4)$
symmetry) and this leads
to the special $1d$ kinematics mentioned. But in our model (with the reduced
$SO(3)$ symmetry), 
lifting the velocity degeneracy causes the energy and momentum conservation to
be distinct constraints in scattering phase space. 
We shall show that in this
case, the reduced (relative to the Luttinger model) scattering cuts off
the X-ray catastrophe
associated with the Luttinger liquid behavior. 
The ``horn-like'' feature in the spectral weight of
the Luttinger liquid is then split into a sharp bound (or anti-bound) state
that co-exists with an incoherent spin-charge decoupled continuum.
We summarize these results in Fig.\ref{schematic}. 

While the main motivation of our model has been to find new
fixed point behaviour in $1d$, 
our model (\ref{themodel})  also has physical  relevance  to two recent
work:
\begin{itemize}
\item  The transport phenomenology of the 
cuprates\cite{2tau} suggests that electrons near the
Fermi surface might divide up into two Majorana modes with 
different scattering rates and dispersion. To date,
this kind of behavior has only been realized in 
impurity models\cite{2CCK} and their infinite dimensional 
generalization.\cite{HoColeman}
We shall show that by breaking the
velocity degeneracy of the original chiral Luttinger model, we obtain
a one-dimensional realization of this behavior: 
a sharp Majorana mode
intimately co-existing with an incoherent continuum of excitations,
reminiscent of the higher dimensional phenomenology. 

\item Frahm et. al.\cite{Frahm} have recently proposed that the low energy
effective Hamiltonian of an integrable spin-1 Heisenberg chain 
doped with mobile spin-$1/2$ holes is 
given by (\ref{themodel}), with one Majorana fermion 
$\Psi^{(0)}$ describing a slow moving
excitation coming from the dopant, interacting with three
rapidly-moving Majorana fermions that describe the spin-1 
excitations\cite{Affleck} of the spin-chain. (See Discussion.) Such doped
spin-chain models may be relevant to certain experimental systems such as
$\rm Y_{2-x} Ca_x Ba Ni O_5$.\cite{spinchain}
\end{itemize} 
 
Whereas the  $SO(4)$ model can be treated
by bosonization,\cite{Haldane,Voit} 
by changing the velocity of a {\it single} Majorana fermion we introduce a
non-linear term into the  bosonized Hamiltonian that preclude
a separation in terms of Gaussian  spin and charge bosons. (See Discussion.)

To tackle this $SO(3)$ model, we have developed a new fermionic
bootstrap method, that has its basis the diagrammatic approach of
Dzyaloshinskii and Larkin (1974)\cite{D&L}. Their method depends crucially
on the existence of conserved currents to eliminate large sets of diagrams,
leading to a closed set of equations that can be solved analytically
for the Green function. On first glance, the reduced number of conserved 
currents in the $SO(3)$ model (compared to the $SO(4)$ model) causes the
Dzyaloshinskii and Larkin method to be inapplicable, because one has to
deal with non-conserved current vertices that involve 
the singlet Majorana fermion
of different velocity. We have found however, that by dealing directly
with fermionic propagators and the four-legs fermionic vertex,
bypassing the intermediate currents, there are enough conservation laws
after all to eliminate all vertex corrections to the skeleton
self-energy  (Fig.\ref{SSE}), allowing us to write down a compact
set of coupled equations involving only the fully renormalized skeleton
self-energy and the exact Green function of the theory. 

The plan of the paper is as follows.
In Section II, we define the class of models of interest here.
In Section III, we describe our modification of the classic 
Dzyaloshinskii and Larkin\cite{D&L} diagrammatic method for
solving one-dimensional fermionic systems, to deal with our case where not
all the velocities are the same. In Section IV, we take advantage
of the purely chiral nature of our model (\ref{themodel}) to
write down a scaling form to  simplifying considerably the 
bootstrap equations derived in Section III. In Section V, we derive
asymptotic solutions for frequencies near the spectral weight singularities,
and demonstrate our results with numerical solutions. In Section VI,
we discuss the nature of this new fixed point.  Some of the 
results have appeared in a brief form in Ho and Coleman\cite{prl}.

\section{MODEL}

The class of model we study here is:
\bea
H = \int dx \Big\{ -i \sum_{a=0}^3 v_a \Psi^{(a)}(x) \partial_x \Psi^{(a)}(x) \cr
+ g \Psi^{(0)}(x) \Psi^{(1)}(x) \Psi^{(2)}(x) \Psi^{(3)}(x) \Big\}   , 
\label{classmodel}
\eea
where $\Psi^{(a)}$ are real (Majorana)  fermions such that 
$\Psi^{(a)}(x)=\Psi^{(a) \dagger}(x)$. The fermions are chiral 
(right-movers, say): this  is a 
crucial property that ensures the system stays gapless, and allows for 
exact solutions in a number of cases. 

In the special case where
all velocities are the same, this model has an
$SO(4)$ symmetry, where the four Majorana
modes can be associated with the spin up and down, electron
and hole excitations of the Fermi surface. To see this, write 
$c_{\uparrow}=\frac{1}{\surd 2} (\Psi^{(1)} - i \Psi^{(2)}),
c_{\downarrow}= - \frac{1}{\surd 2} (\Psi^{(3)} + i \Psi^{(0)})$,
where $c_{\alpha}$ are the usual Dirac fermions, 
and the $SO(4)$ model is just the conventional 
one-branch spin-$1/2$ Luttinger model:
\bea
H_{SO(4)} & = &  \int dx \Big\{ \sum_{a, \sigma} 
c^{\dagger}_{\sigma}(x) i\: v \: \partial_x c_{\sigma}(x) + H.c. \nonumber \\
 & - & g  \left( c^{\dagger}_{\uparrow}(x) c_{\uparrow}(x) - 1/2 \right) 
\left( c^{\dagger}_{\downarrow}(x) c_{\downarrow}(x) - 1/2 \right) \Big\} .
\eea This $SO(4)$ model can be shown by bosonization to be a Luttinger 
Liquid.\cite{Voit}

We shall mostly focus on the $SO(3)$ model where $v_1=v_2=v_3=v \ne v_0$:
\bea
H & = & \int dx \Big\{ H_{kin}(x) 
 + g \Psi^{(0)}(x) \Psi^{(1)}(x) \Psi^{(2)}(x) \Psi^{(3)}(x) \Big\}  , \cr
H_{kin}(x) & = & -i v  \sum_{a=1}^3  \Psi^{(a)}(x) \partial_x \Psi^{(a)}(x) 
-i v_0  \Psi^{(0)}(x) \partial_x \Psi^{(0)}(x) \label{themodel} \cr
  &   &        
\eea Note that this model reduces to the single-impurity
model of Coleman et. al.\cite{2CCK} when the mode $\Psi^{(0)}$ is made 
to localize at the impurity site, and Ho and Coleman have studied the 
same lattice $SO(3)$ model in high dimensions.\cite{HoColeman}
We will show that, 
by making the velocity of one Majorana fermion different,
the scattering phase space decreases drastically, leading to this singlet  
splitting off from the Luttinger continuum to form a sharp 
bound-state/anti-bound-state. 
Thus this is a system that has two qualitatively distinct 
relaxation rates, a dramatic departure from the Luttinger Liquid scenario.

The $SO(2)\times SO(2)$ model where $v_0=v_1 \ne v_2=v_3$ is also
solvable by bosonization, and interestingly, our bootstrap
method also works here. (See the sections Results and  Discussion.)

Finally, we shall also briefly look at the $SO(2)$ model where
$v_0\ne v_1 \ne v_2=v_3$. While we do not know if our method
works here, we expect that due to the separate energy and momentum
conservation, there is still a restriction of scattering phase space, and
the theme of split-off sharp bound/anti-bound-states continues.

Note that the number of  degrees  of freedom and the interaction
are the same in all the cases; 
the variety of behavior seen is due solely to changes in the scattering 
phase space, when the velocities of the fermions are made to be different.

\section{METHOD:PHILOSOPHY}

\begin{figure}
\unitlength1.0cm
\begin{center}
\begin{picture}(4,1.3)
\epsfxsize=6.0cm
\epsfysize=4.0cm
\epsfbox{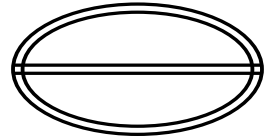}
\end{picture}
\vskip 0.1 truein
\caption{Renormalized ``skeleton self-energy'' (SSE), 
where double lines represent full propagators.
\label{SSE}}
\end{center}
\end{figure} 

Our approach is based on the observation that for the $SO(4)$ and $SO(3)$
models (and possibly others too), the  
renormalized skeleton self-energy,(SSE) containing
full propagators, but no vertex corrections,(Fig.\ref{SSE})
is {\it exact}, so that 
\bea
\Sigma_{a}(x,\tau) = g^2 G_{b}(x,\tau) G_{c}(x,\tau) G_{d}(x,\tau) , \label{nca}
\eea
where the $G_{a}$ are the exact, interacting Greens functions and $\{a,b,c,d\}$ is
a cyclic permutation of $\{0,1,2,3\}$.
These equations close with the usual relations:
\bea
\Sigma_{a}(k,\omega) = (i\omega - v_a k) - 
G_{a} (k,\omega)^{-1}  .  \quad (a=0,1,2,3) \label{dyson}
\eea
Equations (\ref{nca},\ref{dyson}) together define a boot-strap method 
to solve the problem. 

To show that there are no vertex corrections to the  
renormalized skeleton self-energy, we first review and then extend 
Dzyaloshinskii and Larkin's method. 
Provided that we have a minimal $SO(3)$ symmetry, then the
three current densities    
$j^a(x) = -i\epsilon_{abc} \Psi^{(b)}(x) \Psi^{(c)}(x)$ ($a,b \in (1,2,3)$) 
are conserved classically. Following 
Dzyaloshinskii and Larkin\cite{D&L,Metzner,Kopietz}, 
since charge and current are the same in a chiral model, 
the continuity equation guarantees that the 
N-point connected current-current correlation functions 
vanish  for $N>2$: (${\bf x}_i = (x_i, \tau_i)$)
\bea
\langle j^a({\bf x}_1)j^a({\bf x}_2) \dots j^a({\bf x}_N) \rangle_C=0, \quad \quad (N>2)
\eea
For the non-interacting system,
this result leads to the ``loop cancellation theorem'': for
 the amplitude associated with 
a closed fermion loop with $N>2$ conserved current insertions,  
the sum over all possible permutations 
of $\{ {\bf x}_i\}$  of the current operators must give  
zero\cite{D&L,Metzner,Kopietz}. In Appendix A, for
illustration, we give a derivation
for the $N=4$ case and also for odd $N$.
Dzyaloshinskii and Larkin used this cancellation to eliminate all diagrams that
contain such closed loops, considerably simplifying the 
vertex function and polarization bubbles.  

\begin{figure}
\epsfxsize=8.0cm
\epsfbox{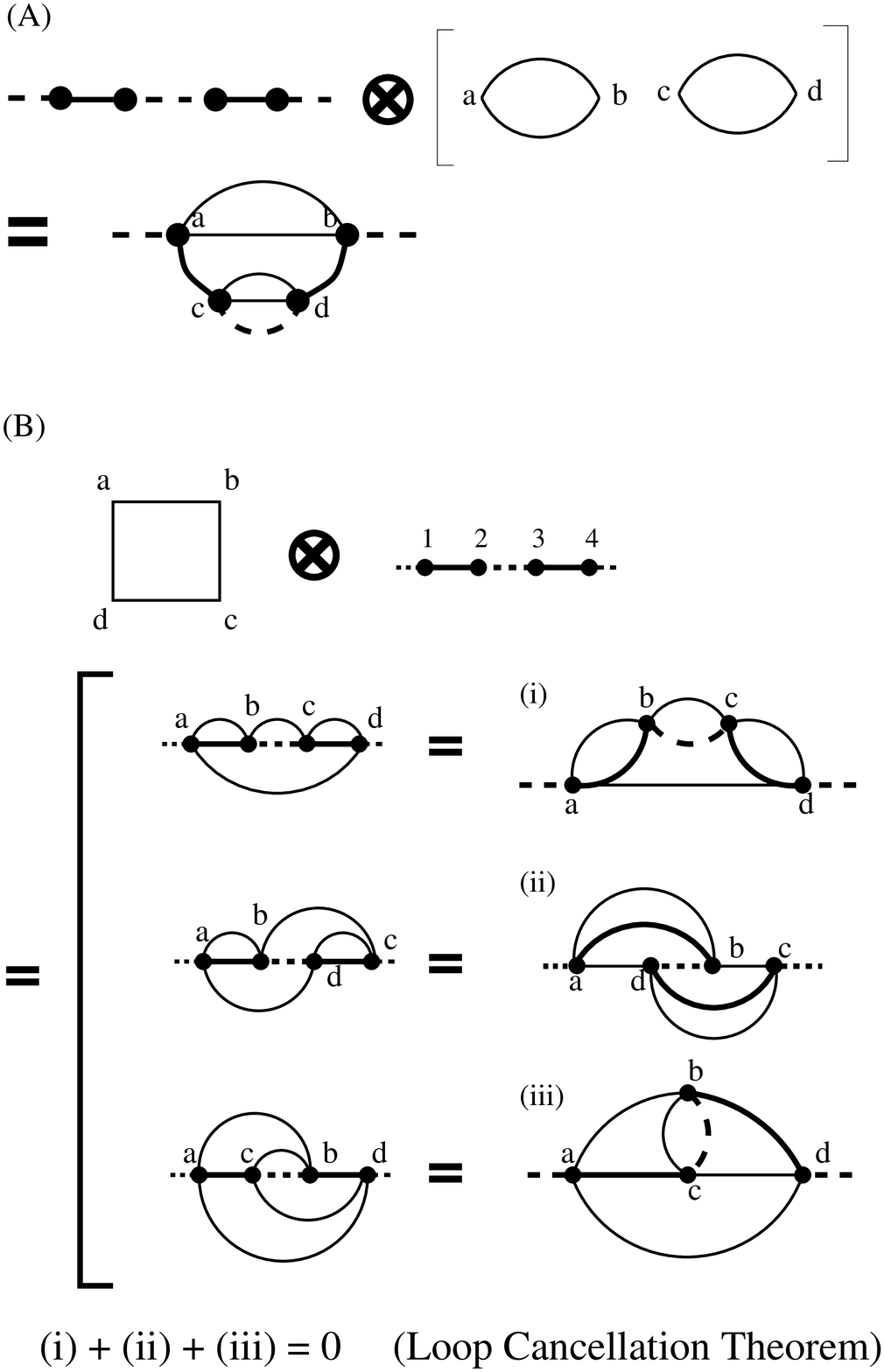}
\vskip 0.1 truein
\caption{(A) Illustrating how the only non-vanishing
singlet self-energy at order $g^4$ is constructed by combining
a propagator back-bone with loops containing  two vertex insertions.
Dotted lines indicate bare propagator for  the singlet
Majorana fermion $\Psi^{(0)}$. Full lines indicate bare propagator
for the triplet Majorana fermions.
(B)  Illustrating how the non-skeleton self-energy 
at order $g^4$ is constructed by combining
a propagator back-bone with loops containing four vertex insertions.}
\label{SSEback}
\end{figure}

We use the loop cancellation theorem in a new way, to show that
the vertex  corrections to the skeleton self-energy (SSE) (Fig.\ref{SSE})
identically vanish.
Unlike Dzyaloshinskii and Larkin\cite{D&L,Metzner,Kopietz}, we discard the
intermediate currents and the associated current vertices, 
and deal only with fermionic propagators and the
four-leg interaction vertex. The Loop Cancellation Theorem is the same.
This method has the advantage that it is more compact 
(only the self-energy and the Green functions are involved), and treats all
propagators in a symmetric manner.
To illustrate the idea, consider the self-energy of the singlet 
Majorana mode in the SO(3) model.
Figs.\ref{SSEback} list all such diagrams 
at order $g^4$.  
The Feynman diagrams contributing
to the skeleton self-energy are  constructed by combining 
loops with two insertions. This is clearly true
for the second order diagram, and we illustrate this using the first
non-trivial order: the 
fourth order diagram in Fig.\ref{SSEback}(A),
but it holds to all orders in perturbation theory. 
Non-skeleton  contributions to the self energy involve 
diagrams with loops containing more than
two current insertions. In these diagrams, the sum 
over all permutations of the current insertions into the
loops is automatically  zero, as illustrated to 
order $g^4$ in Fig.\ref{SSEback}(B). A convenient way to represent these
diagrams is to split each diagram into a backbone
which is the same in all three diagrams, and the 4-insertions loop. Inserting
the 4 vertices of the 4-loop in various ways into the 4 vertices of the
backbone  gives the three diagrams in Fig.\ref{SSEback}(B).
Note that this method of generating the diagrams give rise to the correct
degeneracy for each of the diagram types (i,ii,iii). 

\begin{figure}
\unitlength1.0cm
\begin{center}
\begin{picture}(9,16)
\epsfxsize=9.0cm
\epsfysize=16.0cm
\epsfbox{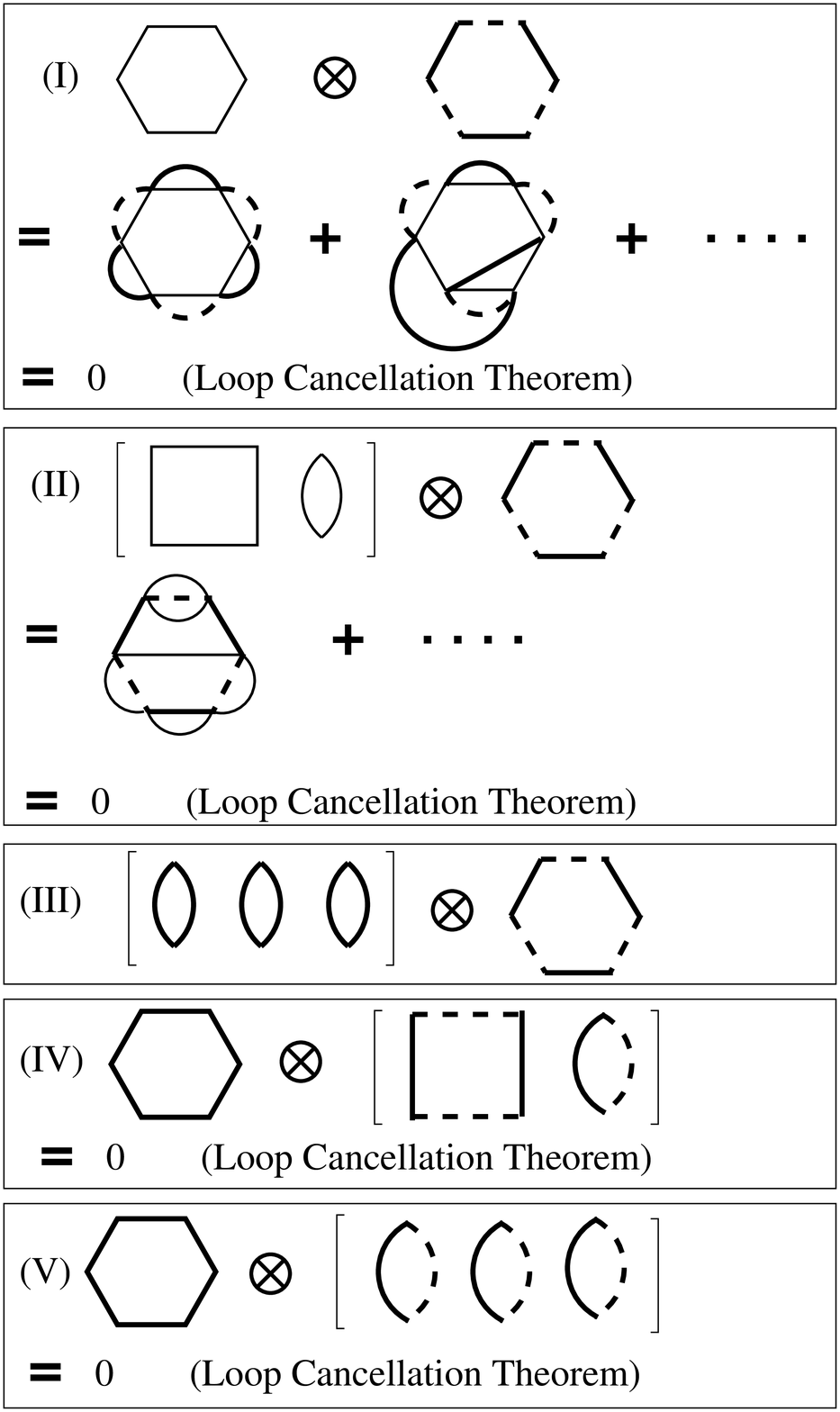}
\end{picture}
\vskip 0.1 truein
\caption{List of all classes of free-energy diagrams that
generate non-SSE diagrams at order $g^6$. 
\label{order6}}
\end{center}
\end{figure}

To generalize these results to higher order graphs, it is more convenient
to look at the set of diagrams for the free-energy. Cutting a $\Psi^{(0)}$
line gives back the singlet self-energy $\Sigma_0$. We first note that
only even
orders in $g$ occur in the free-energy expansion, because
the bare Majorana propagators are diagonal in the Majorana flavor index.
Next, there is always a closed loop with $n$ propagators (not necessary
of the same type) in any of the
free-energy diagrams of order $g^n$. Otherwise, 
improper and/or disconnected self-energy diagrams would be generated. 
Then, at order $g^6$ for example, we have the following classes of
diagrams listed in Fig.\ref{order6} that might generate non-SSE diagrams. 

The Loop Cancellation Theorem applies to each case where there is
a closed loop with more than 2 propagators of the same kind. Thus case
(III) is the only one left. Yet, case (III) generates either SSE diagrams,
improper self-energy diagrams (where cutting one of the lines lead to 2 
disconnected parts), or else, diagrams that have already been
counted in the other cases. The last observation follows from the fact
one can always find a closed 6-loop or 4-loop buried in the diagram.
Hence, all potential non-SSE generating diagrams disappear! One can 
clearly generalize the same reasoning to higher order diagrams. We only
need to check  that this method deals with the combinatoric
factors correctly, i.e, all the degeneracies of the diagrams are
such that there are no non-SSE diagrams left over. 
Here, we appeal to
the fact that in the $SO(4)$ model, there must also be the correct
loop-cancellations, because our method gives the same exact
answer as Dzyaloshinskii and Larkin's method. Even though we
have drawn the diagrams treating the triplet lines as identical,
these triplet lines actually must carry a Majorana flavor index, 
and to generate all
possible diagrams whether distinct under $SO(3)$ or not, we must
draw all possible diagrams with proper indexing of each
of the lines. Listing all diagrams this way is independent of
which symmetry we are dealing with, and consequently, combinators factors
will automatically be taken care of in doing loop cancellation
with these Majorana indices on the propagator lines.
In particular, the symmetry or combinatoric factors for
each diagram must be just right to allow loop cancellation to work in the $SO(4)$
case, and hence for the $SO(3)$ case too.

\begin{figure}
\unitlength1.0cm
\begin{center}
\begin{picture}(7,5)
\epsfxsize=7.0cm
\epsfysize=5.0cm
\epsfbox{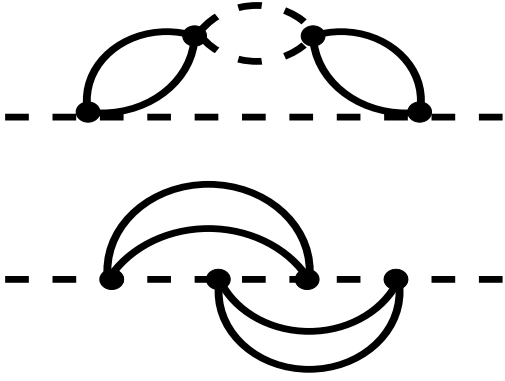}
\end{picture}
\vskip 0.1 truein
\caption{Examples of non-skeleton self-energy
diagrams in the  $SO(2)\times SO(2)$ model
where the Loop cancellation Theorem does not apply.
\label{nocancel}}
\end{center}
\end{figure}

Thus we can show that the vertex corrections to the 
self-energy $\Sigma_0$ of the  singlet Majorana fermion 
cancel to all orders, leaving the fully renormalized   SSE 
as the only remaining contribution. Intriguingly, this argument
fails for the $SO(2)\times SO(2)$ model, because each vertex has
two ``fast'' legs and two ``slow'' legs, unlike in the $SO(3)$ case
where there are only one of the singlet leg. Thus, for example, the
non-SSE diagrams in Fig.\ref{nocancel} do not have a closed loop
of only one kind of propagator to allow loop cancellation to apply.
Yet, these diagrams cannot contribute to the exact self-energy either,
because  the $SO(2)\times SO(2)$ model can be solved exactly by bosonization,
or by a slight extension of Dzyaloshinskii and Larkin's method, and these
results agree exactly with our bootstrap method (See Discussion). 
There must then be more
cancellation than due to just the Loop Cancellation Theorem in its current form.

To complete our proof, we need to show that the triplet
Majorana self-energy is also given by the skeleton diagram.
We  use the {\it full} Kadanoff-Baym Free energy Functional:
\bea 
F[G] = - T \{{\rm Tr} \: {\rm ln} [G^{-1}] + {\rm Tr}[ \Sigma G ] \} + Y[G] ,
\eea
where $Y[G]$ is the sum of all skeleton diagrams.\cite{kadanoffbaym}
Now, by construction, $\delta F[G]/\delta G_a =0$ 
generates the equations for the self-energies,
and in particular, $\delta F[G]/\delta G_0$ must generate
the skeleton self-energy $\Sigma_0$. This requires that the
the Kadanoff-Baym Free energy Functional 
{\sl truncates} at the leading skeleton diagram:
\bea
F = - T \{{\rm Tr} \: {\rm ln} [G^{-1}] + {\rm Tr}[ \Sigma G ] \} +
\parbox{1.0cm}{\epsfig{file=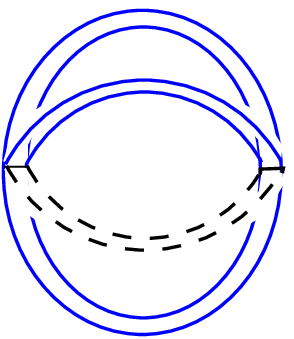,width=1cm,height=1.5cm}}
\eea
Finally, by differentiating the Free-energy functional with respect to the
exact Greens functions $G_{1,2,3}$  of the triplet Majorana fermions, 
each triplet self-energy is also given by the corresponding skeleton self-energy. 

\section{METHOD-DETAILS}

We now apply this result, using the limiting
case of the $SO(4)$ model to check the validity of our results. 
Our equations are dramatically  simplified by seeking 
solutions to (\ref{nca}) which satisfy a  scaling form
\bea
G_{a}(x,\tau) = \frac{1}{2 \pi i x}{\cal G}_{a}(\tau/ ix) , \label{scale}
\eea
This  form is motivated by the observation that 
chirality prevents space from acquiring an anomalous dimension. 
Under a  Fourier Transform, this scaling form is self dual,
\bea
\frac{1}{2 \pi i x} {\cal G}_{a}(\tau /i x) 
\stackrel{F.T.}{\longleftrightarrow} 
\frac{1}{i \omega}{\cal G}_{a}(k/i\omega), \label{ft}
\eea
where the {\it same} function $ {\cal G}_{a}$ appears on both sides.
Inserting Eqn.\ref{ft} into \ref{dyson}  and Fourier transforming,
\bea
\Sigma_{a}(x,\tau) = - \frac{1}{2\pi i x} \frac{d^2}{du^2} 
\left[ 1-v_a u - 1/{\cal G}_{a} (u) \right]_{u=\tau/ix}  .  \label{dysonxt}
\eea
Since the bare Green function scaling form is 
$1/{\cal G}_{a}^0 (u) = 1- v_a u $, it does
not contribute to the self-energy. Combining Eqns. \ref{nca} and \ref{dysonxt},
\bea
\frac{d^2}{du^2} [{\cal G}_{a} (u)]^{-1} = 
- (g/2 \pi)^2 {\cal G}_{b} (u) {\cal G}_{c} (u) {\cal G}_{d} (u)  \label{ncadiff}
\eea
where $\{a,b,c,d\}$ are cyclic permutations of $\{0,1,2,3\}$.  The boundary 
conditions are:
\bea
{\cal G}_{a} (0) = 1, \qquad {\cal G}'_{a} (0) = v_a , \label{ncadiffbc}
\eea
derived from the physical requirement that at high frequencies, the fermions
are free particles, moving with the {\it bare} velocity $v_a$. 
Equations \ref{ncadiff} and \ref{ncadiffbc} are the scaling form version
of our bootstrap method equations \ref{nca}, \ref{dyson}. Notice that
the differential equation \ref{ncadiff}, like Eqn.\ref{nca}, is
independent of the sign of the coupling $g$. Also, (\ref{ncadiff})
has no information on which model of the class (\ref{classmodel}) it
refers to, the symmetry of the model (ie. the velocities) only
come in through the boundary conditions (\ref{ncadiffbc}).

For the $SO(4)$ model,
where ${\cal G}_a(u) \equiv {\cal G}(u)$ $(a=0,\ldots ,3)$, 
Eqns. \ref{ncadiff} reduce to a single differential equation:
\bea
\frac{d^2}{du^2} [{\cal G} (u)]^{-1} = 
- (g/2 \pi)^2 [{\cal G} (u)]^{3}  ,  \label{ncadiffSO4}
\eea  
for which the solution satisfying the boundary conditions ${\cal G}(0) = 1, \quad 
{\cal G}' (0) = v $ is:
\bea
G(x,\tau)= \frac{1}{2\pi i x} \left[1- v_+ \tau/ix \right]^{-1/2} 
\left[1-v_- \tau/ix \right]^{-1/2} 
\label{O4result}
\eea where $v_{\pm}= v \pm (g/2\pi)$ and $v$ is the bare velocity.
Identical results are obtained by bosonization\cite{Voit},
where $v_+$ and $v_-$ are in fact the velocity of the spin-boson and the charge-boson.
Thus this
confirms that the skeleton self-energy is exact for the Luttinger model.

\section{Results}

\begin{figure}
\unitlength1.0cm
\begin{center}
\begin{picture}(7,5)
\epsfxsize=7.0cm
\epsfbox{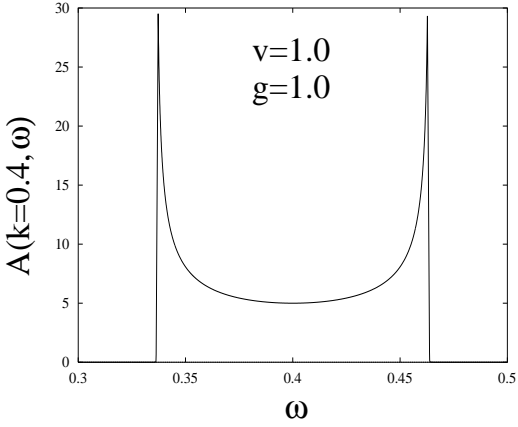}
\end{picture}
\vskip 0.1 truein
\caption{Spectral weight of SO(4) model \label{SO4}}
\end{center}
\end{figure}

\begin{figure}[t]
\unitlength1.0cm
\begin{center}
\begin{picture}(9,6.5)
\epsfxsize=9.0cm
\epsfbox{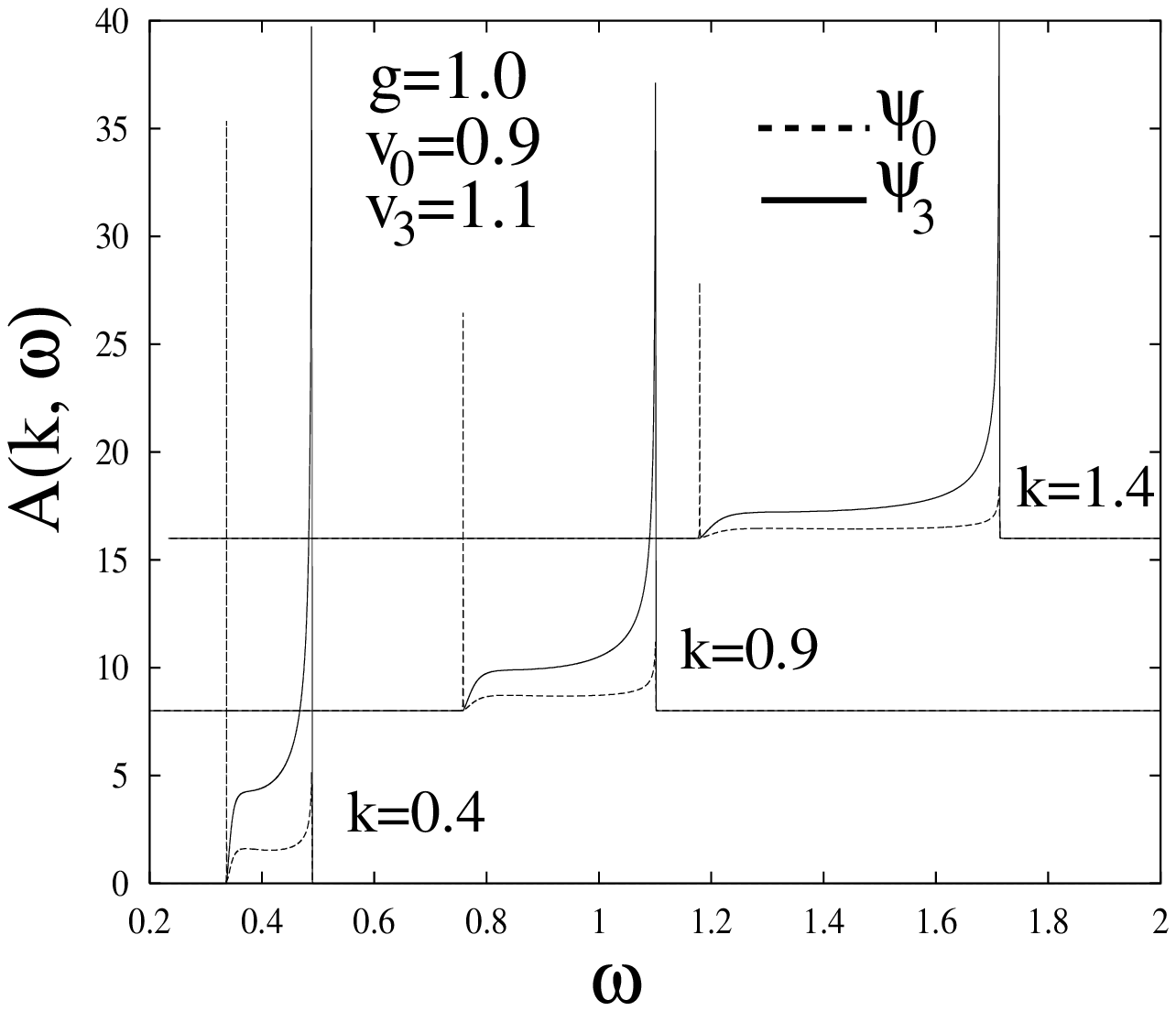}
\end{picture}
\vskip 0.1 truein
\caption{Spectral weight of the SO(3) model. For clarity, we have shifted up
the curves for various momenta by 8 units. \label{SO3}}
\end{center}
\end{figure}

In the $SO(4)$ model, the electron spectral weight displays two 
classic X-ray singularities associated with the decay
of the electron into a  spinon and holon continuum. (Fig.\ref{SO4}).\cite{Voit}
We now show that if $\Delta v = v-v_0$ is finite, one of
these X-ray edge singularities 
is completely eliminated. 
If  $v_0 < v$, we find that low velocity  ``horn'', originally with velocity $v_-$,
develops a 
sharp bound-state pole in the singlet channel, and a broad incoherent excitation
in the triplet
channel with a lifetime growing linearly in energy.
If $v_0> v$, the high   velocity ``horn'' splits off a singlet  anti-bound-state
and the triplet channel develops a high-velocity incoherent excitation. 
(Fig.\ref{SO3}).  The sharp bound-state in the singlet channel develops once a 
velocity difference is introduced, because energy  and momentum
conservation now provide distinct constraints to scattering (unlike in the $SO(4)$ model),
leading to much less phase space for  $\Psi^{(0)}$ to decay into.

\begin{figure}[t]
\unitlength1.0cm
\begin{center}
\begin{picture}(7,5)
\epsfxsize=7.0cm
\epsfysize=5cm
\epsfbox{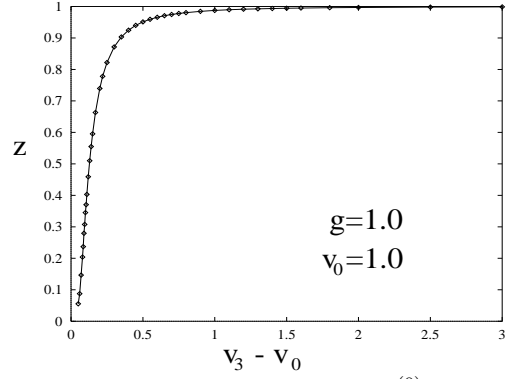}
\end{picture}
\caption{Quasi-particle weight $Z$ of $\Psi^{(0)}$ in the $SO(3)$ model. 
\label{Z}}
\end{center}
\end{figure}

To see this, we must analyze Eqn.\ref{ncadiff} for the $SO(3)$ case:
\bea
\frac{d^2}{du^2} {\cal G} _3^{-1} &=& - (g/2 \pi)^2 ({\cal G}_3)^2 {\cal G}_0,\cr
\frac{d^2}{du^2} {\cal G} _0^{-1} &=& - (g/2 \pi)^2 ({\cal G}_3)^3. \label{diffSO3}
\eea
A very convenient way to discuss these equations is to map them onto a
central force problem. 
If we write ${\bf r} = ( {\cal G} _3^{-1}, {\cal G} _0^{-1})$, 
${\bf F} = - (g {\cal G}_3/{2 \pi})^2(
 {\cal G}_0,  {\cal G}_3)$,
then, $\ddot{{\bf r}} = {\bf F}$,
where $\ddot{{\bf r}} \equiv \frac{d^2 {\bf r}}{du^2}$,
ie., $u$ is like ``time''. By inspection, ${\bf r} \times{\bf F}=0$, 
so the force is radial, thus
the ``angular momentum'', 
${\bf r} \times {\dot{{\bf r}}} = \Delta v$ is a constant. 
If we use polar 
co-ordinates, $({\cal G} _3^{-1},{\cal G} _0^{-1}) = r (\cos \theta,
\sin \theta )$ the equations for the Green-function resemble 
the motion of a fictitious particle 
under the influence of an anisotropic central force: 
\bea
\ddot r - \frac{\Delta v^2 }{r^3}&= &- (g/2 \pi)^2 \frac{1}{r^3 \cos^3  \theta 
\sin \theta}\cr
r^2 \dot \theta &=& \Delta v . \label{central}
\eea
The  velocity difference $\Delta v= v- v_0 $ provides a repulsive
centrifugal force. The boundary conditions (\ref{ncadiffbc}) mean that   
the ``particle'' starts out at $r(0)= \sqrt{2}, \theta(0)= \pi/4$, and with
a slope change $ \dot \theta(0)= \Delta v /2$.

\begin{figure}
\unitlength1.0cm
\begin{center}
\begin{picture}(9,7)
\epsfxsize=9.0cm
\epsfysize=7cm
\epsfbox{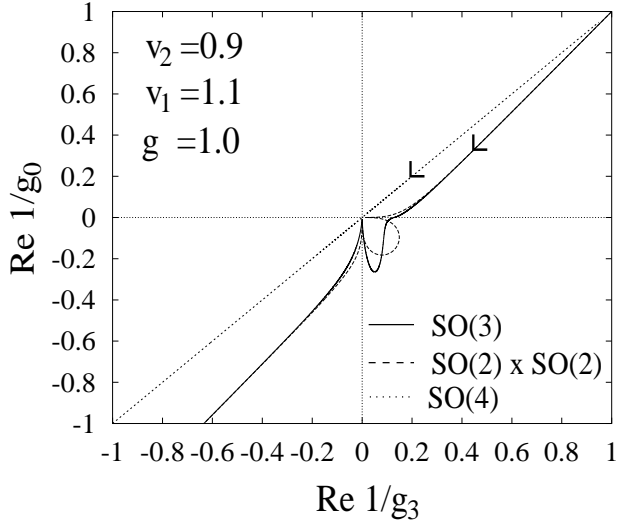}
\end{picture}
\vskip 0.1 truein
\caption{``Trajectory'' in the $(Re 1/{\cal G}_3 ,Re 1/{\cal G}_0)$ plane.
The arrows indicate the direction of increasing ``time'' $u$.
 \label{traject}}
\end{center}
\end{figure}

Without loss of generality, let $\Delta v \leq 0$. For $\Delta v >0$ simply 
replace $v_+\rightarrow v_-$, $g\rightarrow -g$.
When $\Delta v=0$, the ``particle''  falls directly into the origin, and
both ${\cal G}_3$ and ${\cal G}_0$ diverge
with X-ray singularities when the particle first hit the origin at ``time''
$u = 1/v_+$. Then the particle goes pure imaginary in both
coordinates, which gives rise to 
the Luttinger continuum in the spectral weight, until the time $u = 1/v_-$
when the particle goes back to the origin, leading to the other
x-ray singularities for both ${\cal G}_3$ and ${\cal G}_0$. From then on,
the particle stays in the real plane. (Fig.\ref{traject})

  However, once $\Delta v < 0$ is finite, $ \dot \theta(0)= \Delta v$ causes
the orbit to miss the origin at $u \sim 1/v_+$. Instead, 
$\theta\rightarrow 0$ at some finite
``time'' $u= 1/v_0^*$ (Fig.\ref{traject}), 
at which $r=C$ and $\dot \theta =  \Delta v /C^2$. 
For $u\sim 1/v_0^*$,  
it follows that $(r,\theta) = (C, \dot \theta (u-1/v_0^*))$, from which we can
read off the following asymptotics: 
\bea
{\cal G}_3(u)^{-1} & \sim & C   ,\\
{\cal G}_0(u)^{-1}& \sim & (1- u v_0^*)/Z,  \qquad Z ={C v_0^*}/{|\Delta v|} .
\eea Thus  the associated X-ray singularity in the spectral function for
both ${\cal G}_3$ and ${\cal G}_0$
is  eliminated, replaced by an anti-bound state for the
singlet ${\cal G}_0$ with spectral weight $Z$, 
moving with velocity
$v_0^*$, splitting off  above the continuum. After this time, ${\bf r}$ is complex
in both coordinates, until eventually at $u=1/v_3^*$, the particle passes through the
origin, giving rise to the remaining x-ray singularity at $u=1/v_3^*$ in
both ${\cal G}_3$ and ${\cal G}_0$. 

The quantity $\zeta= (v-v_0)/g$ plays the role of a coupling
constant, and approximate  analytic solutions are possible in the limiting cases
of small and large $\zeta$.
For $|\Delta v| >> \frac{|g|}{2 \pi}$
interactions can be ignored, so
$v_0^*\rightarrow  v_0$, and $Z\rightarrow 1^-$.  For $|\Delta v| << \frac{|g|}{2 \pi}$,
the ``motion'' of the fictitious particle  emulates that of the $SO(4)$ model until the
angle $\theta$ approaches zero. We may estimate $v_0^*$ and $C$ by integrating 
Eqn.\ref{central}
with the approximation $r(u) \approx \tilde r(u)$, where 
$\tilde r = [2 ( 1- v_+ u )(1-v_-u)]^{1/2}$ is the $SO(4)$ solution: 
\bea
\int_0^{1/v_0^*} \frac{d\theta}{du} du =  -\frac{\pi}{4} 
& = & \int_0^{1/v_0^*}\frac{\Delta v}{\tilde r^2(u)}du,  \cr
C  & \approx   & \tilde r(1/v_0^*)  .
\eea
After doing the integral, this estimate gives (for $|\Delta v| << |g|/2 \pi$):
\bea
v_0^*  &=&v_+ + \frac{g}{\pi} \exp - \left| \frac{g}{2  \Delta  v }\right|, \\
Z &=& \left|\frac{\sqrt{2} g}{\pi \Delta v }\right| 
      \exp -\left| \frac{g}{4 \Delta v }\right|
\eea
indicating that the formation of the sharp anti-bound-state is non-perturbative
in the velocity difference. 

To illustrate these results further, we have carried out numerical 
solutions of the differential equations (\ref{diffSO3})
for intermediate values of the coupling constant $\zeta$,
using a standard adaptive integration routine.\cite{numrec} 
Results are summarized  in Figs.\ref{SO3},\ref{Z}.

While we have not established the validity of our method to models
of lower symmetry (but see Discussion), 
we believe that the method captures the essence of
the kinematic constraints imposed by energy and momentum conservation,
at least for weak coupling. Thus, we have also performed numerical
calculations for the $SO(2)\times SO(2)$ and  the $SO(2)$ models.

\begin{figure}
\unitlength1.0cm
\begin{center}
\begin{picture}(9,5.5)
\epsfxsize=9.0cm
\epsfysize=5.5cm
\epsfbox{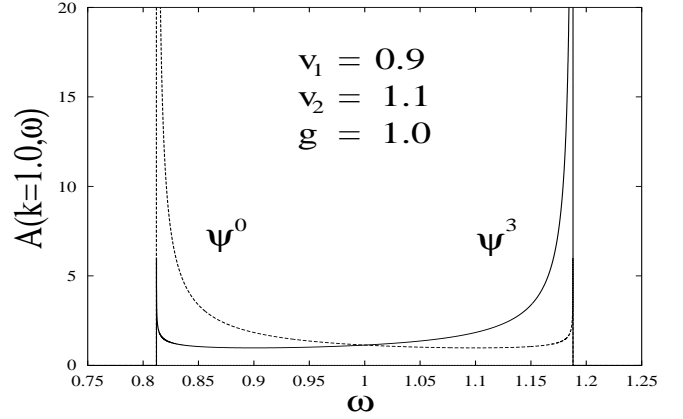}
\end{picture}
\vskip 0.1 truein
\caption{Spectral weight of the $SO(2) \times SO(2)$ model. 
 \label{SO2SO2}}
\end{center}
\end{figure}

For the $SO(2)\times SO(2)$ model, the pair $\Psi^{(0)},\Psi^{(1)}$ 
with the same bare velocity
can combine together to form a boson, and similarly for  $\Psi^{(2)},\Psi^{(3)}$. 
This  leads back to a Luttinger liquid form, but
with asymmetric power law singularities
at the renormalized velocities $v_+$ and $v_-$ (Fig.\ref{SO2SO2}). 
(Also see Eqn.\ref{O2O2} later in Discussion, for the exact analytical solution
for this model.)

\begin{figure}
\unitlength1.0cm
\begin{center}
\begin{picture}(9,6.5)
\epsfxsize=9.0cm
\epsfysize=6.5cm
\epsfbox{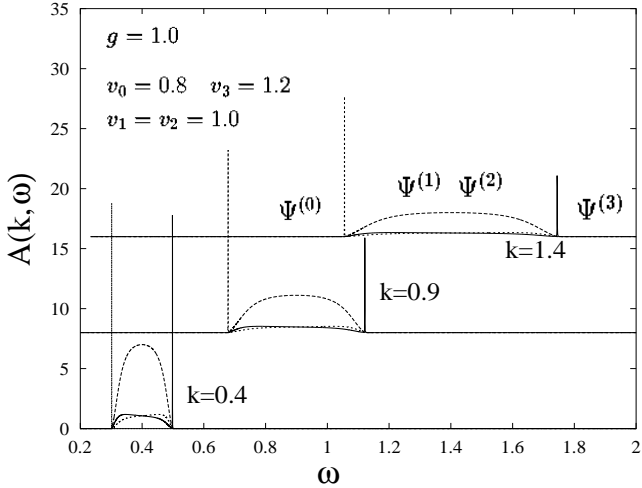}
\end{picture}
\vskip 0.1 truein
\caption{Spectral weight of the $SO(2)$ model. For clarity, we have shifted up
the curves for various momenta by 8 units. \label{SO2}}
\end{center}
\end{figure}

As we progress to the $SO(2)$ case, when $v_0 < v_1 = v_2 < v_3$, we see
a sharp pole for the fermion which has the extremal velocity different to
all the others, while the Luttinger continuum turns into wide peaks linear in energy
for the fermion(s) with intermediate velocities, see Fig.\ref{SO2}. This illustrates
once more our contention that making one Majorana degree of freedom to have a
different (extremal) velocity causes drastic collapse of the scattering phase
space for this fermion.

\section{DISCUSSION and CONCLUSION}

\subsection{The $1d$ Majorana $SO(3)$ Model}

In summary, we have demonstrated that by breaking the velocity
degeneracy of a system of interacting chiral fermions we restrict the
scattering phase space in a way which causes a sharp bound or
anti-bound state to split off from the spin-charge continuum, leading
to a system with two qualitatively distinct spectral peaks and
scattering rates. This is a significant departure from the Luttinger
liquid scenario and demonstrates a new class of  one-dimensional
fixed point behavior.

This new fixed point exhibits properties in common with both 
Luttinger and Fermi liquids, and is perhaps closest in character to the
Marginal Fermi liquid phenomenology introduced in the
context of cuprate metals.\cite{Varma} Like the Fermi liquid,
there is a  sharp quasiparticle bound-state, 
but this co-exists with a Luttinger liquid-like continuum
which is bounded by two extremal velocities.

As mentioned, the $SO(3)$ model cannot be solved by conventional
bosonization, forcing
us to introduce this new bootstrap method. Two immediate questions arise:
\begin{itemize}
\item the nature of the $SO(3)$ fixed point, and
\item the range of validity of the bootstrap method.
\end{itemize}

In the $SO(4)$ model, the fermionic spectral weight has
X-ray singularities at the velocities $v_+ , v_-$ (see Methods-Details).
By bosonization, the model can be mapped onto a theory of
{\it free} bosons (the spin-boson and charge-boson) moving at $v_+ , v_-$,
where for $g>0$, $v_{spin}=v_-$, $v_{charge}=v_+$, and for $g<0$, the
role of $v_+ , v_-$ are swapped round. This is a direct consequence of
separate charge and spin conservation in the model.\cite{Metzner}
We can demonstrate this in the Majorana fermionic representation.
The classically conserved densities are:
\bea 
J_{01}(x) & \equiv & -i \Psi^{(0)}(x) \Psi^{(1)}(x) , \cr
J_{23}(x) & \equiv & -i \Psi^{(2)}(x) \Psi^{(3)}(x) .  \label{Jdef}
\eea (By the $SO(4)$ symmetry, we can define other combinations also.)
Using the commutation relations listed in Appendix B, we get
the equations of motion:
\bea
(-\partial_{\tau} -v q) J_{01}(q) & = & \frac{g}{2 \pi} q J_{23}(q) ,\cr
(-\partial_{\tau} -v q) J_{23}(q) & = & \frac{g}{2 \pi} q J_{01}(q) . \label{EOM}
\eea The right hand side of the equations is not zero (as would be
expected for conserved currents) because of the anomalous commutator (Appendix B):
\bea
[ J_{01}(p) , J_{01}(q) ] = [ J_{23}(p) , J_{23}(q) ] = p \delta (p+q), 
\eea which is the $SU(2)$ level 2 Kac-Moody algebra anomaly.\cite{Tsvelik}
Fortunately, by diagonalizing the system (\ref{EOM}), 
the linear combinations $J_{-}(q) = J_{01}(q) - J_{23}(q)$
and $J_{+}(q) = J_{01}(q) + J_{23}(q)$ do satisfy the continuity equation:
\bea
(-\partial_{\tau} -v_+ q) J_{+}(q) & = & 0 ,\cr
(-\partial_{\tau} -v_- q) J_{-}(q) & = & 0
\eea where $v_{\pm}$ is as before in Eqn.(\ref{O4result}), indicating that these new
densities $J_{\pm}$ are proportional to the spin-boson and charge-boson.\cite{translate}
This then leads to sharp poles in the charge and spin susceptibilities.

For the $SO(3)$ model, using the same definitions (\ref{Jdef}),
we find:
\bea
(-\partial_{\tau} -v_0 q) J_{01}(q) & = & \frac{g}{2 \pi} q J_{23}(q) 
+ (v_3 -v_0) K_{01}(q) ,\cr
(-\partial_{\tau} -v_3 q) J_{23}(q) & = & \frac{g}{2 \pi} q J_{01}(q) . \label{EOM3}
\eea where $K_{01}(q) \equiv -i \sum_k k \Psi^{(0)}(q-k) \Psi^{(1)}(k)$. This
extra term came from the commutator of $J_{01}$ and the kinetic energy,
and causes the set (\ref{EOM3}) not to close, and bosonization in terms of free
spin and charge-bosons (or any linear combinations) is impossible. 
In short, because of the anomaly,
the classically conserved $SO(3)$ density $J_{23}$ is admixed with the
classically non-conserved $J_{01}$, leading to the loss of a sharp pole
for the susceptibility corresponding to $J_{23}$. Thus it is  unlikely that
the model can be mapped to a model of free bosons, which makes it very different
to the conformally invariant fixed points of the $SO(4)$ model and
the $SO(2) \times SO(2)$ model (see below). Also the presence of a sharp
pole in the fermionic spectral weight indicates that there is at least
one fermionic degree of freedom in the diagonalized system. 

Now, Frahm at.al.\cite{Frahm} has conjectured that this $SO(3)$ model
is the low-energy effective theory of an integrable model
of a spin-1 chain doped with spin-$1/2$ mobile holes. Using
Thermodynamic Bethe Ansatz, they have shown that the spin and charge sectors
of the doped holes become decoupled at low temperatures, and
have calculated the low temperature free energy of the spin contribution to be:
\bea
F_{spin} & = &-\frac{\pi T^2}{6 v_0} \left( \frac{1}{2} - \frac{3 A}{4 \pi} \ln A \right) \cr
	 &   &-\frac{\pi T^2}{6 v} \left( \frac{3}{2} + \frac{3 A}{4 \pi} \ln A \right) 
+ \ldots  .\label{freeenergy}
\eea where $A>0$ is a constant that depends on the doping only. With $A=0$ (undoped case),
the first term has been interpreted\cite{Frahm} as coming 
from a single Majorana fermion of
velocity $v_0$, while the second term come from a triplet of 
massless Majorana fermions with velocity $v$
that represent the $SU(2)$ level 2 WZNW model, which has been
shown by Affleck\cite{Affleck} to be the low energy effective theory of
the gapless integrable spin-1 chain. Naively, one expects that a system of fermions with
two velocities cannot be conformally invariant, unless the two species do not interact
with each other and thus form two decoupled sectors that are {\it individually}
conformally invariant.\cite{cfi} If the free energy (\ref{freeenergy}) is indeed the 
$SO(3)$ model low temperature free energy, the form of the free energy suggests
that the $SO(3)$ model is again conformally invariant asymptotically, and hence
the diagonalized (presumably fermionic) basis consists of two decoupled sectors. 
Note that the diagonalized basis is unlikely to coincide with the bare Majorana
fermions of Eqn.(\ref{themodel}), because the coupling is marginal, at
least up to $O(g^3)$.\cite{unpubl}

As for the range of validity of the bootstrap method we introduced to solve the model,
we note that
if we change two Majorana velocities at the same time,
so that $v_0=v_1$ and $v_2=v_3$, we would have reduced the symmetry still
further, to an $SO(2) \times SO(2)$ symmetry.\cite{Penc} We can solve the
differential equations (\ref{ncadiff}) with the results
\bea
G_{3}(x,\tau)&=& \frac{1}{2\pi i x} 
\left[1-\frac{v_+ \tau}{ix} \right]^{-\frac{1}{2}+\gamma} 
\left[1-\frac{v_- \tau}{ix} \right]^{-\frac{1}{2}-\gamma},\cr
v_{\pm}&=& \frac{1}{2}\bigl[v_0 + v_3 \pm [
(v_3 -v_0)^2 + (g/\pi)^2]^{\frac{1}{2}} \bigr],  \label{O2O2}
\eea 
and 
$\gamma = \frac{1}{2}(v_3 -v_0)\{{(v_3 -v_0)^2 + (g/\pi)^2}\}^{- \frac{1}{2}}$. 
Interestingly, this model can be bosonized to a model of free bosons,
and the bosonization result agrees exactly with (\ref{O2O2}). 
This is surprising because
as far as we can see, the closed-loop cancellation 
is not sufficient in the case of the $SO(2)\times SO(2)$ model to cancel all
vertex corrections. This suggests
that a more general cancellation principle is at
work, and that the range of validity of our solution may even extend
to models with a still smaller, $SO(2)$ symmetry. To date, 
we have not been able to prove this result.  

We also wish to point out
that our differential version (\ref{ncadiff}) of the bootstrap equations
(\ref{nca},\ref{dyson}) are of such a simple form only because we have
a purely chiral system. If we allow Left- and Right- movers to interact,
 the scaling form (\ref{scale}) no longer applies\cite{Voit}, and we have
not found a different scaling form that allows similar simplifications.
However, we expect the bootstrap method to work still, as long as
there are separate conservation of Left- and Right- currents. This is 
true at least for the $SO(4)$ model, because Dzyaloshinskii and Larkin\cite{D&L}
have shown that their method works also for such systems, and our method
is a generalization of theirs.

\subsection{Broader Issues: higher dimensions?}

\begin{figure}
\unitlength1.0cm
\begin{center}
\begin{picture}(9,6.5)
\epsfxsize=9.0cm
\epsfysize=6.5cm
\epsfbox{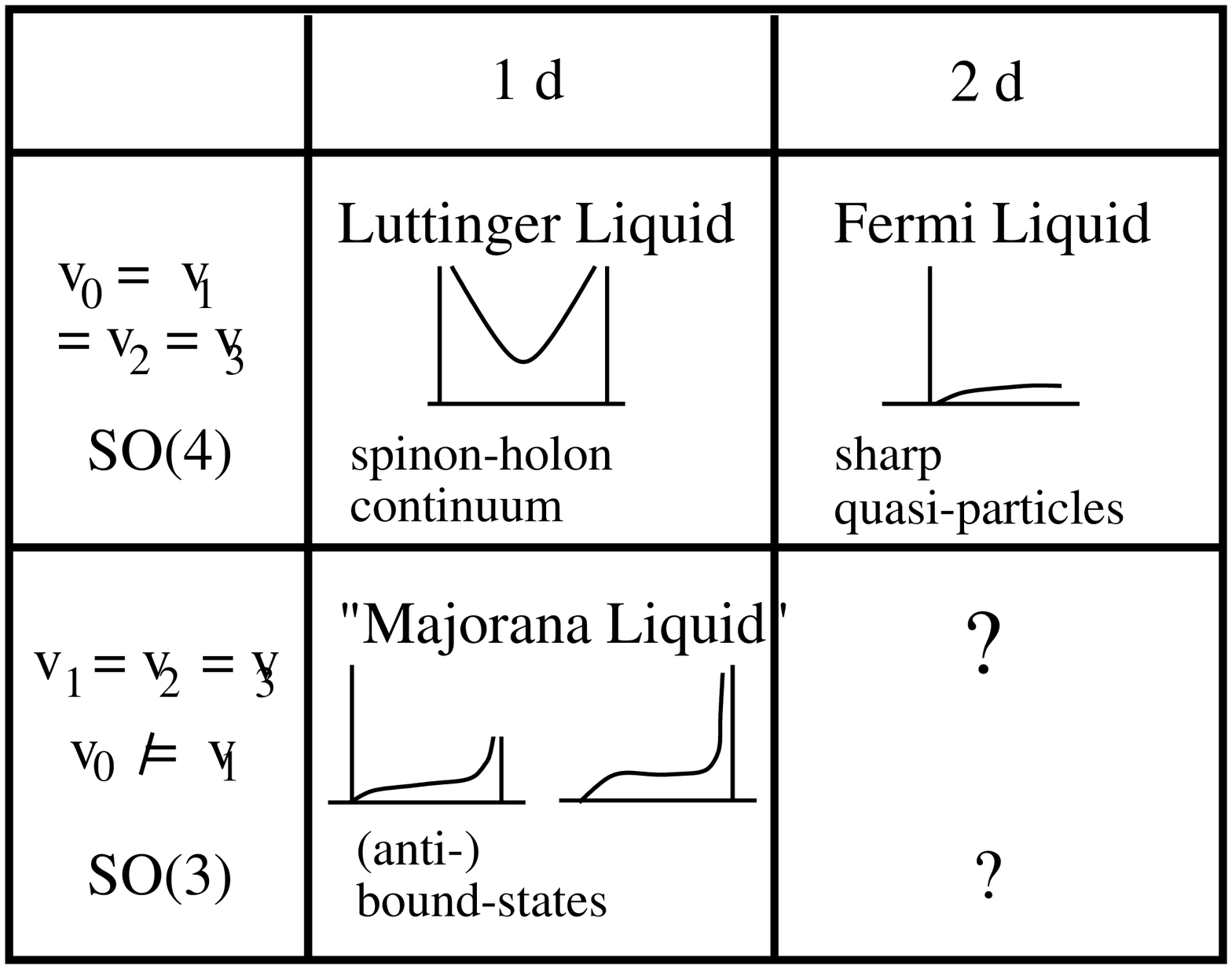}
\end{picture}
\vskip 0.1 truein
\caption{ \label{table}}
\end{center}
\end{figure}

Our work raises the question whether this kind of non-Fermi Liquid
behavior might survive in dimensions higher than one.  In higher
dimensions energy conservation and momentum conservation are distinct
constraints on scattering phase space, and the Luttinger liquid
reverts to a Fermi liquid, at least for short-range
interactions\cite{Metzner,Shankar}.  By contrast, the SO(3) model
cannot be solved by bosonization, and its unusual properties have
reduced reliance on the special kinematics in $1d$.
 Thus, this kind of
behavior might be more robust in higher dimensions. In fact, near
infinite dimensions\cite{HoColeman}, two lifetimes behavior persists
in the $SO(3)$ model, but here, the thermodynamics near zero
temperature is that of a Fermi Liquid. The case of small, but finite
dimensions is however, still open.

{\em Acknowledgment}
We should like to thank Natan Andrei, Thierry Giamarchi, Alexei Tsvelik,
Achim Rosch, Edmond Orignac, Revaz Ramazashvili, Andrei Lopatin
and particularly Walter Metzner for discussions related to this work. 
This work was supported by NSF grant NSF DMR 96-14999.

\section{Appendix A}

In this appendix, we prove by diagrammatic method,
the Loop Cancellation Theorem for
a loop with four current insertions. It is easiest to
prove this in $x,\tau$ space.
(For a proof in momentum-frequency space, see Kopietz et.al.\cite{Kopietz}.)
Let the four insertions be at ${\bf x}_i = (x_i, \tau_i)$, $i=1,\ldots,4$.
Each leg of the loop is a free propagator:
\bea
G_{ij} \equiv G(x_i, \tau_i;x_j, \tau_j) = \frac{1}{v(\tau_i-\tau_j) + i (x_i-x_j)}.
\label{bareG}
\eea
 Denote by $[1234]$ the loop where
going clockwise starting from  ${\bf x}_1$, we encounter successively
${\bf x}_1,{\bf x}_2,{\bf x}_3,{\bf x}_4$, ie.,
\bea
[1234] =  G_{43} G_{32} G_{21} G_{14}.
\eea
Without loss of generality, we can fix ${\bf x}_1$ and sum over 
permutations of the other three vertices.
The Loop Cancellation Theorem then says:
\bea
[1234] + [1243] + [1342] + [1324] + [1423] + [1432] = 0
\eea
But for even number of propagators in a loop, going clockwise is
the same as going anti-clockwise, hence, e.g. $[1243]=[1342]$. So,
we only need to prove:
\bea
[1234] + [1243] + [1324] = 0 .
\eea
To do this, we need the important identity:
\bea
G_{ij} G_{jk} = G_{ik} (G_{ij} + G_{jk}) , \label{id}
\eea
which can be proven simply by substituting in Eqn.\ref{bareG}.
Use this to rewrite the loops:
\bea
[1234] & = & G_{14} G_{43} G_{32} G_{21} =  G_{13} (G_{14} +G_{43}) G_{32} G_{21} \cr
[1243] & = & G_{13} G_{34} G_{42} G_{21} =  G_{13} (G_{34} +G_{42}) G_{32} G_{21} \cr
[1324] & = & G_{14} G_{42} G_{23} G_{31} =  G_{31} (G_{14} +G_{42}) G_{12} G_{23}
\eea and it is clear that they do all cancel, since $G_{ij} = - G_{ji}$.

>From this example, we can see that it is important for the cancellation 
of loops with even number of current insertions,
that all the propagators be of the same type, to use the identity (\ref{id}).
In our context, this means all the propagators are for fermions of the
same velocity. 

For an odd number of insertions, the identity (\ref{id})
is not needed, because time-reversal invariance guarantees the cancellation:
a loop $[1ijkl\ldots xyz]$ will be cancelled by the counter-clockwise partner
$[1zyx\ldots lkji]$, thanks to $G_{ij} = - G_{ji}$ and a total of odd
number of propagators.
(This is the analogue of Furry's theorem in QED, see eg. 
Peskin and Schroeder\cite{Peskin}.)

We note in passing that this identity (\ref{id}) can also be used to prove the
Ward Identity by diagrammatic methods, order by order.\cite{Peskin}
(For further information on the Ward Identity and how it is used for
diagrammatic methods for finding the exact Green Function in
some one-dimensional systems, see Metzner et.al.\cite{Metzner}.)

\section{Appendix B} 

Here we list some commutation relations used to derive the equations
of motion of the various (classically) conserved densities. Start from
the canonical anti-commutation relation for the Majorana fermions:
\bea
\{ \Psi^{(a)}(x) , \Psi^{(b)}(y) \} = \delta_{ab} \delta(x-y) .
\eea With the definitions (\ref{Jdef}), and with the $SO(3)$ Hamiltonian
$H=H_0 + H_{int}$:
\bea
H_0 & = &\int dx  -i \sum_{a=1}^3 v \Psi^{(a)}(x) \partial_x \Psi^{(a)}(x) \cr
    &   & -i v_0  \Psi^{(0)}(x) \partial_x \Psi^{(0)}(x) ,  \cr
H_{int} & = & -g \int d x J_{01}(x) J_{23}(x)
\eea we can get straightforwardly:
\bea
[J_{01}(p), H_0] & = & v p J_{01}(p) + (v-v_0) K_{01}(p) ,\cr
[J_{23}(p), H_0] & = & v p J_{23}(p) .
\eea We can recover $SO(4)$ results by setting $v=v_0$. Ordinarily,
we would expect $[J_{01}(p), H_{int}]=0$, but this is spoilt by
the $SU(2)$ level 2 anomalous commutator\cite{Tsvelik}:
\bea
[ J_{01}(p) , J_{01}(q) ] = [ J_{23}(p) , J_{23}(q) ] = p \delta (p+q) .
\eea One can derive this by eg. a diagrammatic method, see Ch. 13 of the
book by Tsvelik\cite{Tsvelik}. This then leads to the only non-trivial
commutation relations:
\bea
[J_{01}(p), H_{int}] & = & \frac{g}{2\pi} p J_{23}(p) ,\cr
[J_{23}(p), H_{int}] & = & \frac{g}{2\pi} p J_{01}(p) .
\eea

\newpage


\begin{references}
\bibitem{translate} Indeed using the definition 
$c_{\uparrow}=\frac{1}{\surd 2} (\Psi^{(0)} - i \Psi^{(1)}),
c_{\downarrow}= - \frac{1}{\surd 2} (\Psi^{(2)} + i \Psi^{(3)})$,
the densities $J_{01} \propto c^{\dagger}_{\uparrow} c_{\uparrow}$,
$J_{23} =  c^{\dagger}_{\downarrow} c_{\downarrow} -1/2 $ are just the spin
up and spin down densities (relative to the mean, half-filled case), 
hence, $J_{01}+J_{23}$ is the total charge
density, and $J_{01}-J_{23}$ is the spin density.
\bibitem{cfi} This is in fact the case for the two bosonizable models mentioned
here: the $SO(4)$ model can be bosonized to a free spin-boson with
velocity $v_{spin}$ and a free charge-boson with $v_{charge}$. 
For the $SO(2)\times SO(2)$ model, where $v_0=v_1$ and $v_2=v_3$, this model
bosonizes to two free bosons again, with velocities  $v_+$ and $v_-$ as 
mentioned in DISCUSSION.
\end{references}
\end{document}